# Study of Band structure, Transport and magnetic properties of $BiFeO_3$-$TbMnO_3$ composite

Prince Kr. Gupta[1], Surajit Ghosh[1], Arkadeb Pal[1], Somnath Roy[2], Amish G Joshi[3], A. K. Ghosh[2] and Sandip Chatterjee[1,*]

[1]Department of Physics, Indian institute of Technology (BHU), Varanasi – 221 005, India

[2]Department of Physics, Banaras Hindu University, Varanasi – 221 005, India

[3]CSIR-National Physical Laboratory, Dr. K. S. Krishnan Road, New Delhi – 110 012, India

*Corresponding Author's e-mail: schatterji.app@iitbhu.ac.in

## Abstract

Magnetoelectric multiferroic composite of two types of multiferroic (Type I and II) consisting $BiFeO_3$ and $TbMnO_3$ is studied for enhanced magnetic and transport properties. A narrower band gap is estimated from the UV-visible absorption spectrum from that of BiFeO3 and $TbMnO_3$. With known value of band gap, the band structure was estimated from the valence band x-ray photoemission spectra (XPS) and ultra violet photoemission spectra (UPS). The valence and conduction band was found at 1.0 eV and 0.45 eV above and below the Fermi level respectively. Thus the insulating behavior of the system is understood from the reconstruction of the energy bands at the interface which happens due to lattice mismatch of the two materials. The large coercivity and the increase on the magnetization value are understood to be due to superexchange interaction between different Mn ions ($Mn^{2+}$, $Mn^{3+}$ and $Mn^{4+}$). From the composition study of EDXA and core level x-ray photoemission spectra oxygen vacancy was found which in turn creates the mixed valence state of Mn to maintain the charge neutrality.

## Introduction

Multiferroic materials have been attracting researchers recently for their interesting fundamentals as well as for the possibility of application of these materials in different spintronic devices [1-4]. In magnetoelectric multiferroic materials ferroic (or anti-ferroic) magnetic and electric ordering co-exist in a single phase giving rise to the possibility of controlling the magnetization (intrinsic polarization) with the application of electric field (magnetic field) [1]. Due to these coupling between the two properties magnetoelectric multiferroic materials have become one of the most important materials of today [1]. The reason of the limited number of multiferroic material is the mutual exclusive origin of the two ordering (empty d shell for ferroelectricity and partially filled d shell for magnetic ordering) [1-4].

$BiFeO_3$ is one of the most interesting and well studied multiferroic as it is the only multiferroic material to show both the ordering (magnetic and ferroelectric) above the room temperature (ferroelectric Curie temperature $T_N \sim 1103$ K and Neel temperature $T_N \sim 643$ K) [5-8]. It exhibits G type canted antiferromagnetic ordering with a cycloid frequency of ~ 62 nm [8]. $BiFeO_3$ shows large spontaneous polarization of order 10-100μc/cm$^2$ because of polar displacement of cations and anions relative to each other pointing along one of the eight pseudo-cubic [111] axes [5-7]. The lone pair electron at the 6s shell of Bi is considered to be the main reason behind the observed ferroelectricity on $BiFeO_3$ while the partially filled 3d shell of Fe is responsible for the canted antiferromagnetic ordering [5-7]. In spite of having these features $BiFeO_3$ is not considered suitable for many applications due to many reasons. Structural instability, difficulty in synthesizing single phase, High leakage current and low resistivity of the material due to presence of $Fe^{3+}$ and oxygen vacancies are some of them [9-12]. To overcome the shortcomings there have been numerous attempts involving doping of different transition metal at Fe site and rare earth metals at Bi sites, inducing chemical pressure or strain in the system [13-16]. An alternate option is to prepare composite of different multiferroic materials with similar structure involving $BiFeO_3$. There are reports on composite structures and superlattice structures of $BiFeO_3$-$BaTiO_3$, $BiFeO_3$-$PbTiO_3$, $BiFeO_3$-$BiMnO_3$ showing improvement in multiferroic properties [17-19]. Yu et al, have reported exchange bias and other enhanced magnetic properties in $BiFeO_3$-$La_{0.7}Sr_{0.3}MnO_3$ heterostructure due to charge transfer assisted band reconstruction near the interface which in turn creates additional ferromagnetic and antiferromagnetic exchange interactions [20].

In this context we have prepared composite of $BiFeO_3$ and $TbMnO_3$, belonging to different type of multiferroic. $TbMnO_3$ is type II multiferroic material in which the ferroelectric ordering ($T_C$ ~ 28 K) arises as a result of magnetic ordering ($T_N$ ~ 42 K) and the material is known to posses strong magnetoelectric coupling [21, 22]. It crystallizes in orthorhombic perovskite (*Pbnm*) close to that of $BiFeO_3$ which has rhombohedral perovskite (*R3c*) structure [21, 23]. The lattice mismatch between them is expected to trigger band reconstruction near the interface [20]. Band reconstruction can influence many physical properties of the composite including transport and magnetic properties. In this report we have studied the band structure through valence band x-ray photoemission spectroscopy (XPS) and Ultraviolet photoemission spectroscopy (UPS) UV-Visible absorption spectroscopy. The effect of band reconstruction on the transport and magnetic properties has also been studied.

**Experimental detail:**

The compounds $BiFeO_3$, $TbMnO_3$ and the composite of $BiFeO_3$ and $TbMnO_3$ (in 7:3 ratio) i.e., 0.7BFO-0.3TMO were prepared following a conventional solid state reaction method by taking the precursors for all the constituent elements i.e., $Bi_2O_3$, $Fe_2O_3$, $Mn_2O_3$ and $Tb_4O_7$ in proper stoichiometric ratio. The detail of the synthesis procedure can be found in other works [24]. The surface morphology and grain growth are studied from field emission scanning electron microscope (FESEM, Nova Nano SEM 450). The composition of the samples was also studied from energy dispersive x-ray analysis (TEAM EDS SYSTEM with Octane Plus SDD Detector) integrated with the SEM. X-ray Photoemission Spectroscopy (XPS) experiments were performed using Omicron Multi-probe® Surface Science System, GmBH, equipped with a dual anode non-monochromatic Mg/Al X-ray source (DAR400), a monochromatic source (XM 1000) and a hemispherical electron energy analyzer (EA 125). All the XPS measurements were performed inside the analysis chamber under base vacuum of ~ $1.8\times10^{-10}$ Torr using monochoromatized $AlK_\alpha$ with a power of 300 Watt. The total energy resolution for monochomatic AlKα line with photon energy 1486.70 eV, estimated from the width of the Fermi edge, was about 0.25 eV. The pass energy for Survey Scan Spectra and core level spectra was kept at 50 eV and 30 eV, respectively. Ultraviolet photoemission spectroscopy (UPS) and UV-Visible absorption spectroscopy were employed to study valence band structure and electronic properties. To investigate the fine changes near Fermi level, Fermi-edge Ultraviolet photoemission spectra were collected using the non-monochromatic He I (21.2 eV) line at an average base pressure of $2.8\times10^{-8}$ Torr. The energy resolution of the analyser and the step size were set at 0.03 eV and 0.05 eV, respectively. The magnetic measurements were carried out in Magnetic Property Measurement System (SQUID-MPMS, Quantum Design, USA). The temperature dependent transport properties were measured in Close Cycle He cryostat (Advanced Research Systems, Inc.) using a Keithley 6517B Electrometer.

## FESEM micrograph and EDX analysis

In order to visualize surface morphology and microstructure of the composite images of sintered 0.7BFO-0.3TMO composite along with the pure BFO and TMO were recorded in the field emission scanning electron microscope (FESEM) [shown in figure 1(a)-(c)]. It can be observed that pure BFO pellet exhibits rectangular and non- homogenously distributed grains with smooth surfaces showing crystal growth of BFO. The average grain size was calculated using image j software with an average size of ~2.16 μm. TMO microstructure is consisted of non-homogenously distributed grains of irregular shape with average grain size of 1.19 μm. The composite also show similar morphology with irregular shaped grains but the grain size varies in broad range as compare to that of BFO and TMO grains. The average grain size also lies in between that of the two pure compounds (1.59 μm) [25]. In order to identify all the elements present in 0.7BFO-0.3TMO composite energy dispersive x-ray analysis (EDXA) analysis was carried out which is shown in figure 1(d). The obtained EDXA result confirmed the homogenous mixing and phase formation of the composite. The EDXA result also indicated to the presence of oxygen vacancy in the system. Manganite systems sintered at higher temperature tend to have oxygen vacancies which in turn can influence the Mn ions present in the system to have multiple valence states in order to maintain the charge neutrality [26].

## Fourier transform infrared (FTIR) spectroscopy

Fourier transform infrared (FTIR) spectroscopy has been one of most used techniques these days to analyze different functional groups and corresponding vibrational bands. The FTIR spectrum of the composite 0.7BFO-0.3TMO was recorded in the wave number range of 400 – 2000 $cm^{-1}$ which is presented in figure 2. The vibrational bands of metal-oxygen bonds generally lie below 1000 $cm^{-1}$. In the case of pure $BiFeO_3$, the bending and stretching vibrations of the Fe-O in $FeO_6$ octahedra and the vibrational band of $BiO_6$ octahedra lie in the range from 400 to 560 $cm^{-1}$ which has been clearly observed in the spectrum (peaks denoted as A and B) [27, 28]. The presence of $TbMnO_3$ perovskite phase in the composite is confirmed by identifying the stretching vibration of Mn-O-Mn bond ~ 573 $cm^{-1}$ arising from $MnO_6$ octahedra of the compound (denoted as C in the spectrum). Apart from these bands few other bands can be identified in the spectrum such as, 1720 $cm^{-1}$ assigned to the bending vibrations of $H_2O$ which arises due to the presence of moistures in the powder [27, 29]. The bands ~ 1070 $cm^{-1}$ and 1440 $cm^{-1}$ are ascribed to the stretching vibrations of C=O and C-C respectively which appears due to absorption of carbon during the exposure of the sample to air [30].

**XPS analysis:**

X-ray photoemission spectroscopy (XPS) is a prominent technique followed by many researchers to study the chemical composition and valence state of constituent elements of a material. We have studied the survey scan and detail scan of Fe2*p*, Mn2*p*, O1*s*, Bi4*f* and Tb3*d* spectra to gain detail knowledge of their electronic structure and composition in the composite $BiFeO_3$-$TbMnO_3$. All the peak positions were matched with the National Institute of Standards and Technology (NIST) XPS database. From the survey it was confirmed that all the constituent elements (i.e., Bi. Fe, Tb, Mn and O) were present in the composite [figure 3(a)]. The absence of any foreign element in the compound was also confirmed from the spectrum. Although, a C 1*s* peak can be seen from the survey scan spectrum, the presence of carbon does not influence any physical property of the composite. In many systems such surface absorbed/adsorbed carbon has been found which does not have any effect on the physical properties. The detail spectra were deconvoluted using the XPSPEAK software. In figure 3(b) the deconvoluted spectrum of the Fe2*p* core level is shown which is split into two broad peaks due to spin orbit coupling. The characteristic peaks of $Fe2p_{3/2}$ and $Fe2p_{1/2}$ are observed around the binding energies 710.6 and 724.7 eV, respectively which correspond to $Fe^{3+}$ state in BFO [31]. In addition to the characteristic peaks there is a satellite peak which is observed at the binding energy position 719 eV. The presence of the satellite peak ~ 719 eV confirms that trivalent oxidation state is the dominant state of iron in BFO which is expected for $BiFeO_3$ [32, 33]. The XPS analysis indicates that only pure $BiFeO_3$ phase is present without any type of impurities.

Figure 3(c) shows the O1*s* XPS spectrum of the composite which is split into two parts revealing two kinds of chemical state of oxygen present in the composite. The peaks at binding energy position ~ 529.3 eV correspond to the oxygen situated at lattice denoted as $O_L$ and the peak positioned at ~ 531.3 eV corresponds to the surface chemisorbed oxygen denoted by $O_V$. The surface chemisorptions of the of oxygen ions occurs due to the presence of oxygen vacancy in the system. In $BiFeO_3$ based compounds it is common to find the oxygen vacancies which are created at the surface due to lattice defects [34]. Moreover, our XPS results are confirms with EDX analysis which indicated the presence of oxygen vacancy in the system.

In figure 3(d) the deconvoluted Mn2*p* spectrum is presented which is also found to be split into two characteristic spin orbit peaks for $Mn2p_{3/2}$ and $Mn2p_{1/2}$ ~ 642.6 and 654.0 eV respectively. Interestingly, the peaks were asymmetric in nature indicating multiple valence state of Mn ions in the composite. The $Mn2p_{3/2}$ and peak was deconvoluted into three characteristic peaks at binding energies 640.4, 641.4 and 642.6 eV which correspond to $Mn^{2+}$ $Mn^{3+}$ and $Mn^{4+}$ respectively. The deconvoluted peak positions for the $Mn^{2+}$ $Mn^{3+}$ and $Mn^{4+}$ species were found at 651.8, 652.8 and 654.0 eV respectively. From the deconvoluted characteristic peaks $Mn^{3+}$ was found to be more intense indicating dominant presence of $Mn^{3+}$ which is obvious in case of $TbMnO_3$. The mixed phase of the Mn ions evolves in the system to maintain the charge neutrality which is affected by the oxygen vacancy observed from the O1s detail spectrum. It has been observed that, in Mn based systems, charge transfer effect between the Mn3*d* orbitals and O2*p* ligand orbitals can give rise to the mixed phase of Mn [35 and references therein]. The creation of $Mn^{4+}$ and $Mn^{2+}$ species can also understood to be due the valence instability of $Mn^{3+}$ ions. The mix valence state of Mn ions is known to modify the bulk magnetic and electrical properties of different manganite systems [36].

In figure 3(e) core level XPS spectra of Tb3*d* element is presented. The spectra shows spin orbit coupling peaks of Tb 3d positioned at ~1276.7 eV and ~1241.8 eV for $3d_{3/2}$ and $3d_{5/2}$ respectively. The peak positions as well as the doublet separation clearly suggest that Tb is in 3+ oxidation state [NIST]. The core level spectra of Bi4*f* is also studied and shown in figure 3(f). The spin orbit split peaks of $Bi4f_{5/2}$ and $Bi4f_{7/2}$ are observed to be present at ~163.7 eV and ~158.3 eV which corroborate well with earlier reports of trivalent valence states of Bi [37, 38].

**UV-Vis Spectroscopy:**

The presence of different charge state of TM ions at the interface due to the charge transfer between Mn-Mn ions may lead to the band reconstruction at the interface. As a result of the reconstruction band gap of the composite is expected to decrease [39, 40]. Thus to confirm the band gap reduction we have measured the absorption spectrum in the UV and visible range. For studying the absorption characteristics of the composite, absorbance at different wavelength (k) (range of 200–800 nm) were recorded and the absorption coefficients (a) were calculated at corresponding wavelengths. As can be seen from the spectrum the absorption band edge lies beyond the range of measurement. We have followed the Tauc's method to estimate the band gap of the composite from the absorption spectra [20]. The photon energy (hν) and the band gap energy for a particular transition are related by the equation;

$$(\alpha h\nu) = K(h\nu - E_g)^n \tag{3}$$

where α is the absorption coefficient given by α = 2.303 (Ab/t), (here, Ab is absorbance and t is thickness of the cuvette which is 1 cm), K is the edge width parameter. The value of n depends on the type of transition, i.e., allowed direct, allowed indirect, forbidden direct, and forbidden indirect for which it can have values 1/2, 2, 3/2, and 3, respectively. $BiFeO_3$ is known to have a bang gap of 2.1 eV to 2.7 eV in different form of the material such as bulk, nanomaterials, or single crystalline material [41-43]. Since $BiFeO_3$ and $TbMnO_3$ both are known to be direct band gap material, the band gap of the composite was determined from the linear fitting of the straight line part of the $(\alpha h\nu)^2$ versus photon energy (hν) plot on the hν axis [44]. From the Tauc's plot (figure 4) it is evident that the band gap of the material lies ~ 01.45 eV which is low compared to other reported value for $BiFeO_3$ [41-43]. Therefore the reduction in the band gap of the composite system confirms the band reconstruction phenomenon due to the charge transfer between the TM ions in the composite.

**Valence band spectra study by XPS and UPS:**

The XPS valence band (VB) spectra of the 0.7BFO-0.3TMO system has been recorded at room temperature (300 K) to gain better insight into its detailed electronic structure (figure 5(b)). It is evident from the figure that near Fermi level ($E_F$=0 eV), the spectral weight of the electronic states is extremely weak or merely absent which essentially suggests for an insulating ground state of the system. Moreover, it is relevant here to mention that the UV-visible spectroscopy study yielded a high band gap of ~1.45 eV and the resistivity data exhibited an insulating nature with high resistance (discussed later). Hence, our VB spectra is corroborating well with our earlier results. The energy position of the valence band maximum has been estimated to be ~1 eV by making a linear extrapolation of the sharply rising feature immediately below the Fermi level (as shown in inset of figure 5(a)) [45]. Apart from this, knowing the band gap of the system (~1.45 eV) and the valence band maximum energy position (~1 eV), the position of the conduction band (CB) minimum can be readily determined to be ~ -0.45 eV. Hence, a schematic diagram of its possible density of states (DOS) (band structure)has been depicted in figure 5(a).

Furthermore, three main features of XPS VB spectra below the $E_F$ can be observed at ~2.5 eV, 7 eV and 10.5 eV which are denoted as $VB_1$, $VB_2$ and $VB_3$ respectively. The complete occupied VB spectra below $E_F$ (ranging from 0 eV to ~14 eV) is mainly composed of hybridized states of Tb4*f* , extended Mn3*d*, Fe3*d* and O2*p* orbitals. Since, for the present system, the Mn and Fe are in octahedral co-ordination with ligand oxygen ions, the crystal field effect causes the Mn/Fe3*d* states to split into $e_g$ and $t_{2g}$ levels. Moreover, for such kind of co-ordination, the $t_{2g}$ states always lie below the $e_g$ states. Therefore, the spectral weight (in the range of 0 eV to ~1 eV) immediately below $E_F$ can be mainly attributed to the contributions from partially occupied Mn3$d_e_g$ and Fe3$d_e_g$ orbitals [46]. However, the first shoulder like feature $VB_1$ positioned at ~2.5 eV has appeared mainly due to the hybridization of Tb4*f*, extended Mn3$d_t_{2g}$, Fe3$d_t_{2g}$ with O 2*p* orbitals. However, significant contribution to this feature $VB_1$ is expected to be from Tb4*f* states as the other states (Mn3$d_t_{2g}$, Fe3$d_t_{2g}$) are extended over a long range [46]. The most intense feature of the VB spectra at ~7 eV ($VB_2$) can be attributed mainly to the hybridization of the Mn3$d_t_{2g}$ and Fe3$d_t_{2g}$ orbitals with the O2*p* states. On the other hand, the feature $VB_3$ at the lowest energy ~10.5 eV emanates from the hybridized states of O2*p* - Mn3$d_t_{2g}$/Fe3$d_t_{2g}$ and other oxygen bonding states O2*p* - Mn/Fe4*sp* and O2*p* - Tb5*sd* etc with significant contribution being from O2*p* states [46].

Moreover, to probe how the VB spectral features get modified with higher resolution and also to augment with our previous results of electronic structure near Fermi level, we have carried out the ultraviolet photoemission spectroscopy (UPS) measurement on the same system (figure 5(c)). It is interesting to note here that the spectral features in UPS VB spectra immediately below the $E_F$ are of relatively low intensity as compared to those observed in XPS VB spectra. This in turn predicts the fact that the region immediately below $E_F$ has significant contribution from the Mn/Fe3*d* states since the photo-ionization cross-section for the Mn/Fe3*d* photoemission is considerably higher than that for the O2*p* photoemission in the XPS process. As a consequence, the low energy UPS is less sensitive to the heavier atoms and highly sensitive to the lighter atoms such as oxygen. Hence, in the whole UPS spectra, the predominant contribution comes from the O2*p* states while the contributions from the Tb4*f*, Mn3*d* and Fe3*d* states get suppressed. Irrespective of the above facts, the UPS VB spectra agreed mostly with the XPS VB spectra. The absence of the electronic states near the Fermi level in the UPS VB spectra supports our previous results, thus suggesting its insulating nature. Moreover, a feature ($VB_1$) near ~2.5 eV (associated to the hybridized states of O2*p* - Mn/Fe3*d* and O2*p* – Tb4*f* orbitals) can be visible in the UPS VB spectra which corroborate with the XPS VB spectra. Similarly, the second feature ($VB_2$) which is the intense broad peak near~7 eV matches well with the most intense peak of the XPS VB spectra. Another weak feature ($VB_3$) near~11.5 eV can also be visible which is seemingly associated to the O*2p*-Mn/Fe*3d* hybridized states with significant contribution coming from O*2p* states. The position of the last feature differs by ~1 eV with that of the $VB_3$ feature observed in XPS VB spectra. This can presumably be attributed to the possible charging effect due to the high resistivity of the system.

**Transport properties**:

Figure 6 shows the resistivity vs. temperature plot of the composite measured in the temperature range 10-300 K. The resistivity of the sample increases continuously with decreasing temperature indicating semiconducting nature of the composite. This also indicates the involvement of activation process in the transport mechanism. Below 42 K the resistivity of the sample increases sharply. The inset (a) of figure 6 shows lnρ vs. 1000/T plot signifying

Arrhenius fitting which shows that our data poorly fit the Arrhenius equation. Hence, the Mott's variable range hopping (VRH) mechanism was considered to govern the transport process. The resistivity in case of VRH model can be written as; [47]

$$\rho = \rho_0 \exp\left[\left(\frac{T_M}{T}\right)^{\frac{1}{4}}\right]$$

Where $T_M$ is knows as Mott's characteristic temperature and is expressed as;

$$T_M = 21.2\ \alpha^3 / k_B N(E_F)$$

The most probable hopping distance (R) and the hopping energy (W) can be presented as;

$$R = \left[{}^{8\pi\alpha N(E_F) k_B T}\!\big/_{9}\right]^{-\frac{1}{4}}$$

$$W = {}^{3}\!\big/_{4\pi R^3 N(E_F)}$$

where, 1/α is the localisation length. The average ionic radii of the two transition metal ions ($Fe^{3+}$ and $Mn^{3+}$) which are considered responsible for the conduction mechanism are taken as the localisation length. Inset (b) of figure 6 shows the lnρ Vs. $(1/T)^{1/4}$ plot which displays a good linear fitting thus supports the transport mechanism to be VRH. It is noteworthy to mention here that in disordered systems with random distribution of defects, VRH mechanism can generally be found to control the transport process [48]. In our sample there are site defects due to presence of mixed valence of Mn ions. The large strain produced at the interface of the two materials due to lattice mismatch would definitely create lattice defects in the composite system. Moreover, the high resistivity of the sample is consistent with band gap ascribed from the UV-Visible spectrum and the band structure calculated from the VB XPS and UPS spectra.

**Magnetic Properties:**

To gain knowledge about the effect of band reconstruction and creation of different valence state of Mn ions on the magnetic property of the composite, we have measured the room temperature magnetization (M) vs. magnetic field (H) hysteresis loop of the composite and compared with that of pure BFO. Figure 7 presents the measured M-H loops for the magnetic field $\pm$7 T from which it can be seen that the magnetization increases linearly with the applied field. This unsaturated behavior is consistent with the antiferromagnetic nature of $BiFeO_3$ at room temperature. A closer look around the zero field reveals that at low fields there is deviation from the linearity and the loop is wide open. This along with the loop opening even at high fields signifies ferromagnetic contribution in the magnetic property of the sample. There are reports on the room temperature weak ferromagnetism in the nanoparticles of $BiFeO_3$ [49]. Also $BiFeO_3$ is known to possess canted antiferromagnetic ordering with a cycloid frequency ~ 62 nm which is governed by the Dzyaloshinskii-Moriya interaction [5, 6, 49]. The weak ferromagnetism can also arise from the canted antiferromagnetic ordering. Inset in the figure 7 shows the M-H loop for the pure BFO sample. It can be seen that the magnetization of BFO also does not saturate showing its antiferromagnetic nature. Interestingly, it can be noticed that the magnetization value increases by an order of two. The coercivity of the material is found to increase from that of pure BFO. The increase and the low field weak ferromagnetic behavior can be understood to be due to the possible ferromagnetic superexchange interaction between different Mn ions ($Mn^{2+}$, $Mn^{3+}$ and $Mn^{4+}$ as seen from XPS spectra) at the interface. According to Goodenough-Kanamori rules the superexchange interactions between $Fe^{3+}$-O-$Mn^{3+}$/$Mn^{4+}$ are also expected to be ferromagnetic [20, 50-51].

## Conclusion

To summarize, we have prepared the $BiFeO_3$-$TbMnO_3$ composite (7:3) via conventional solid state reaction and studied the interface through different charaterizations. Morphological detail and the grain growth were studied from the SEM images. Chemical state and the composition of the material were studied from XPS and EDXA which revealed oxygen vacancy is there which in turn creates mix valence state of Mn ions to maintain the charge neutrality. A remarkable decrease in band gap was observed from the UV-visible absorption spectra from that of $BiFeO_3$. Based on the band gap (~ 1.45 eV), results from XPS valence band spectra and UPS spectra the band structure of the material was drawn in which the conduction band edge was found ~ 0.45 eV. In the valence band three main features were observed at binding energy positions ~2.5 eV, 7 eV and 10.5 eV ($VB_1$, $VB_2$, $VB_3$ and $VB_4$) which were composed of hybridized states of Tb*4f* , extended Mn*3d*, Fe*3d* and O*2p* orbitals. The most intense feature ($VB_2$) were attributed to the hybridization of the Mn$3d_t_{2g}$ and Fe$3d_t_{2g}$ orbitals with the O*2p* states while the weak shoulder like feature $VB_1$ which close to the Fermi energy is attributed to the Tb4*f* states. Such band diagram and the reduction in the band graph are understood to due to the reconstruction of the bands due to interfacial strain. Moreover, the transport property was found to be dominated by variable range hopping mechanism and the high resistivity of the material was also found to be consistent with the band diagram. Antiferromagnetic like non saturating M-H loop were observed with weak ferromagnetic feature at low fields which was attributed to the superexchange interaction between different Mn ions.

**Acknowledgement:**

Authors would like to acknowledge the central instrument facility centre of IIT (BHU) for SEM, EDXA and magnetic measurements.

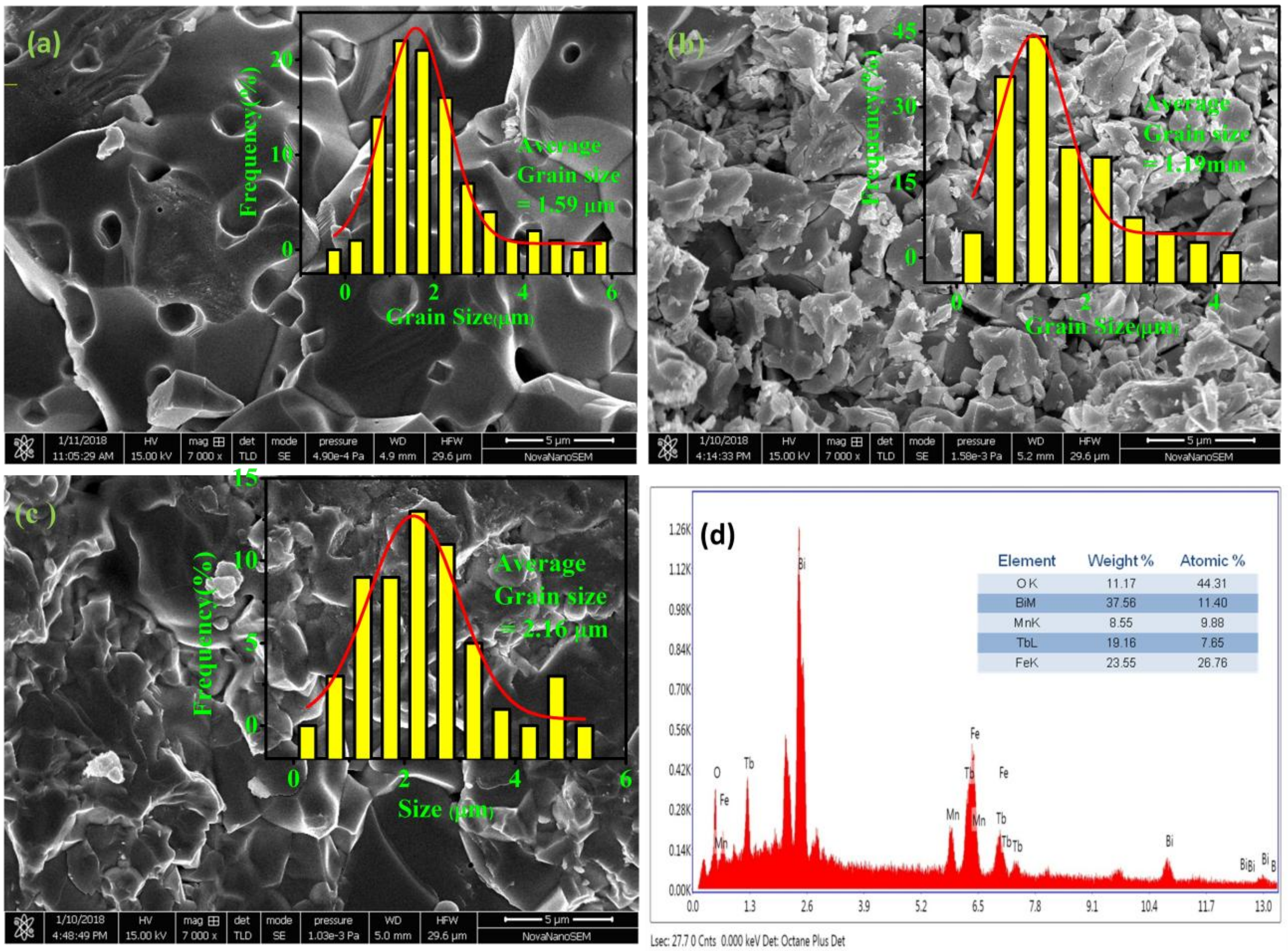

**Figure 1 (a)-(c):** FESEM images from fractured surfaces of the single phase BFO, TMO and composite 0.7BFO-0.3TMO respectively and insets shows the corresponding calculation of grains size bye image j software. **(d)** shows EDXA graph of the composite 0.7BFO-0.3TMO with its chemical composition in the insets.

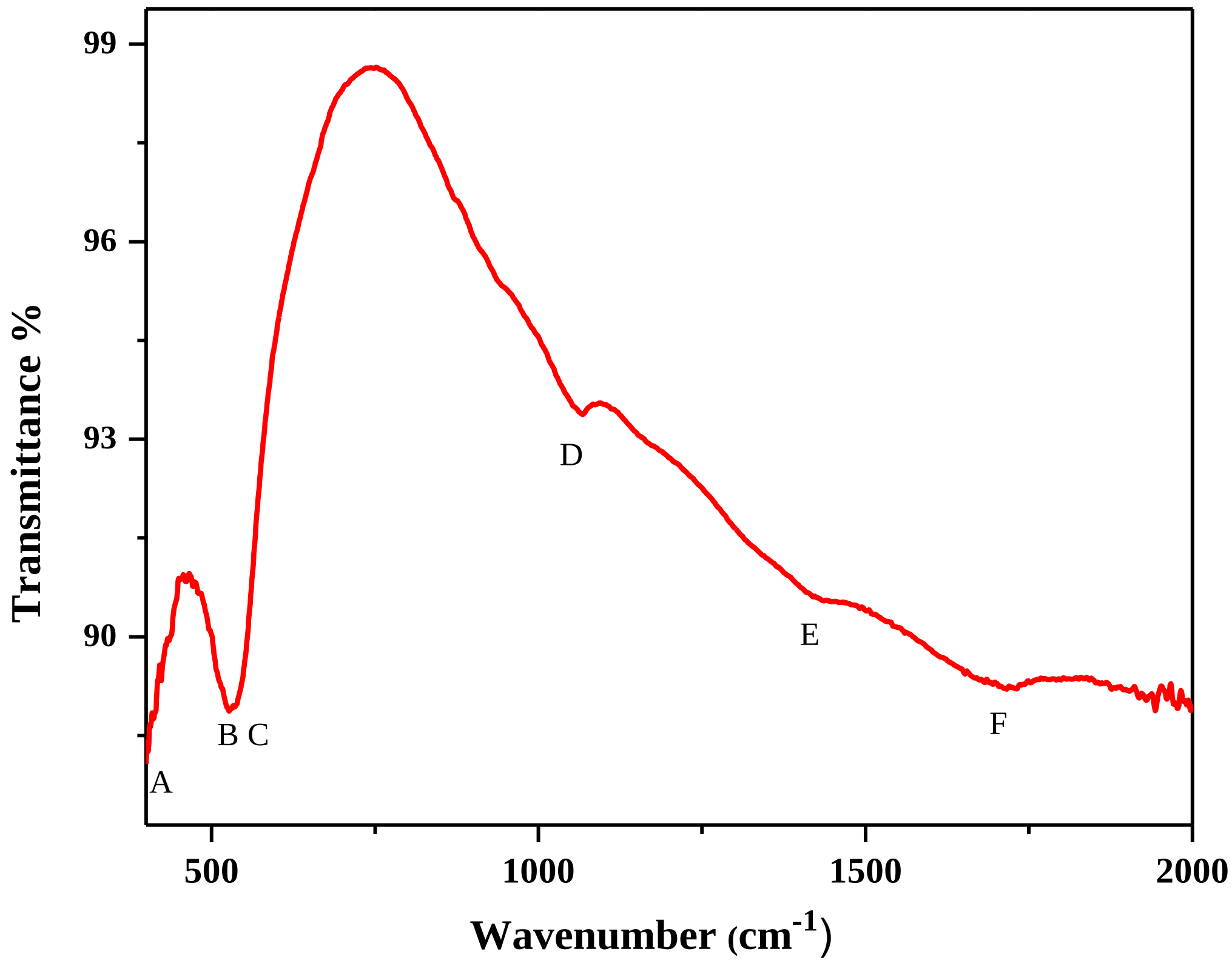

**Figure 2**: FTIR spectrum of the composite 0.7BFO-0.3TMO recorded at 300 K.

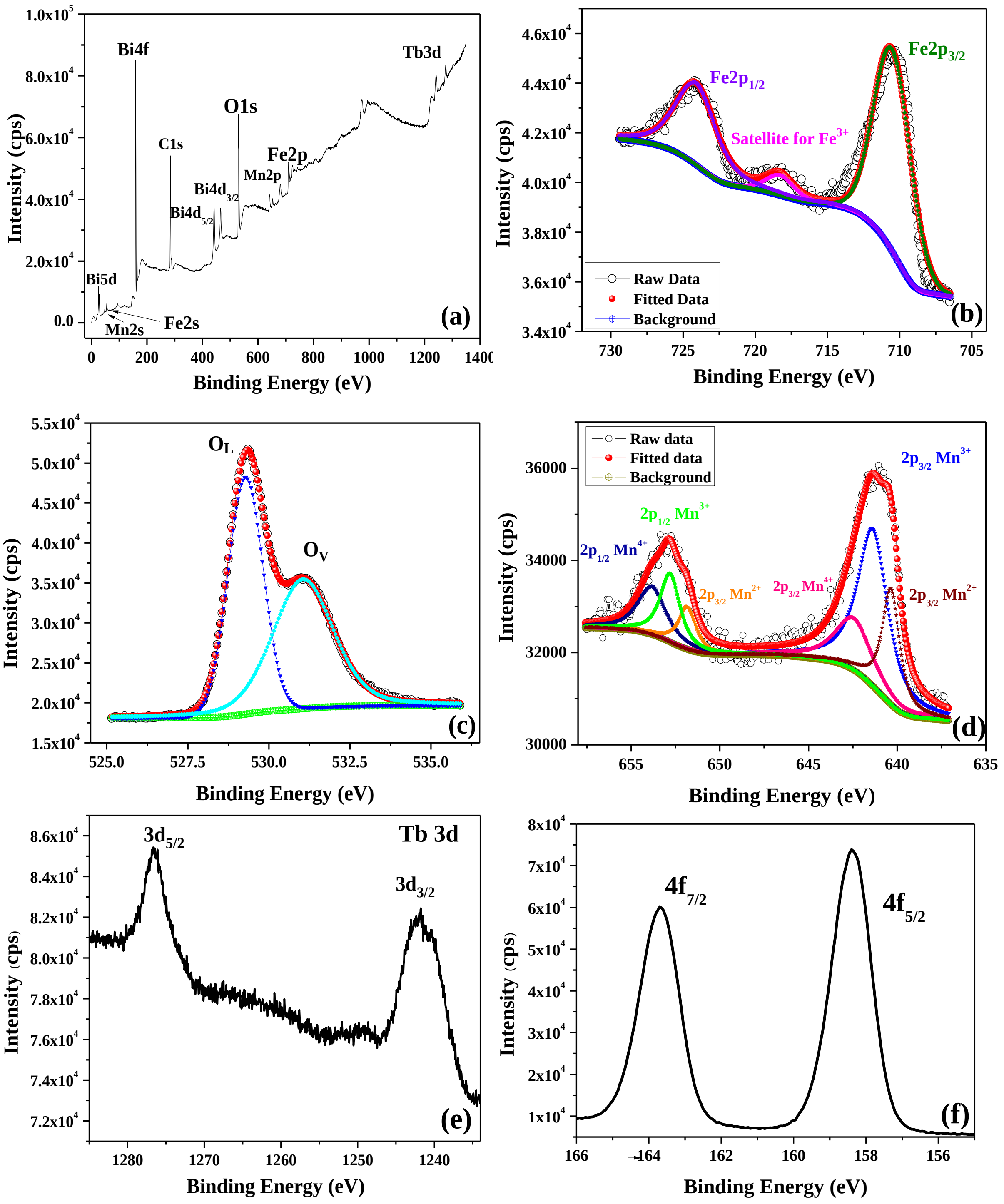

**Fig. 3: (a)** XPS survey scan of 0.7BFO-0.3TMO at 300 K. Deconvoluted detail spectra of **(b)** Fe 2*p* **(c)** O 1*s* **(d)** Mn 2*p* of composite 0.7BFO-0.3TMO. Detail spectra of **(e)** Tb 3*d* and **(f)** Bi 4*f*.

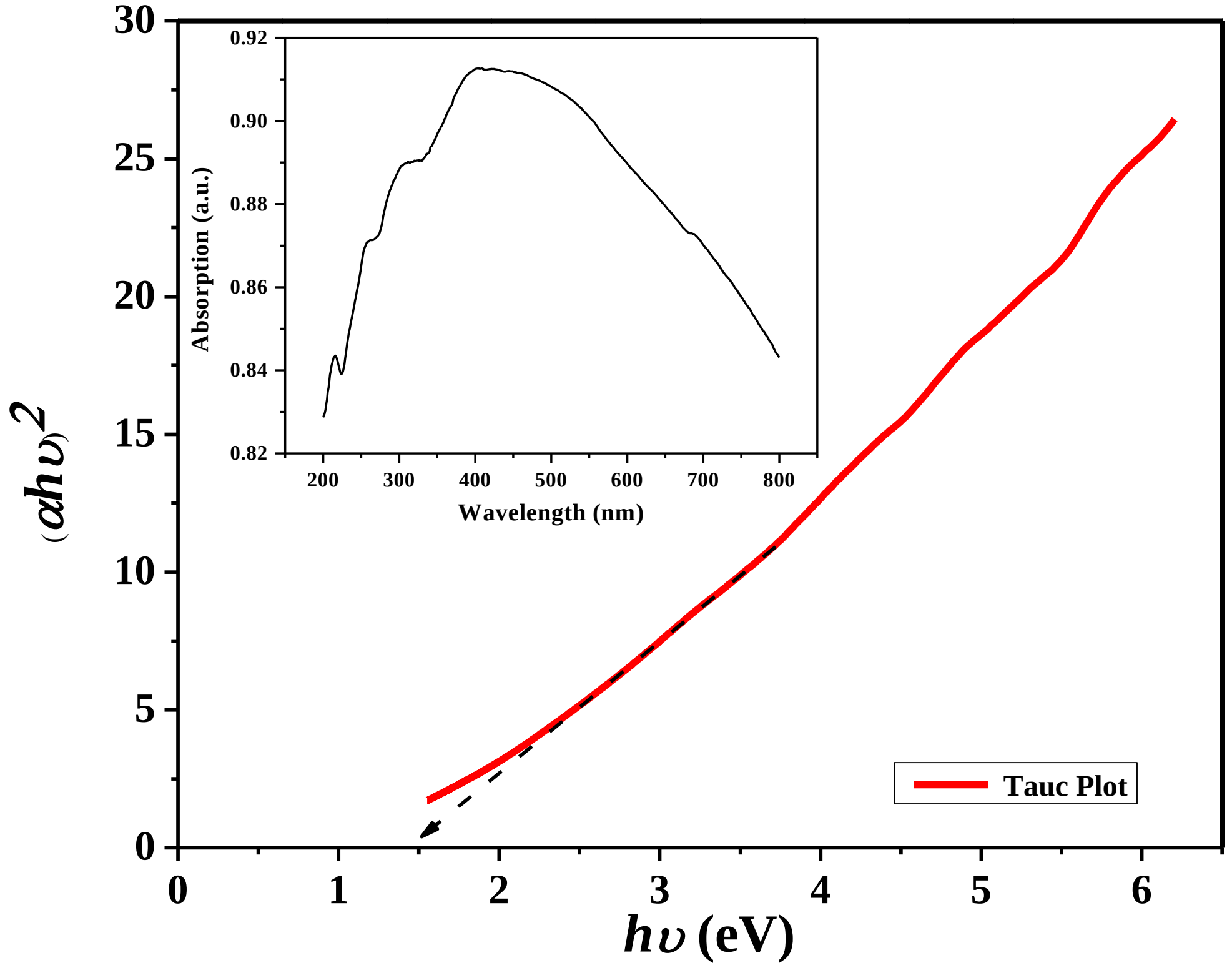

**Figure 4**: Tauc plot for the determination of the optical band gap of the composite 0.7BFO-0.3TMO at room temperature. The black dashed arrow is guide to the eye, showing the extracted bad gap. The inset shows the absorption spectrum of the composite from which the Tauc plot has been estimated.

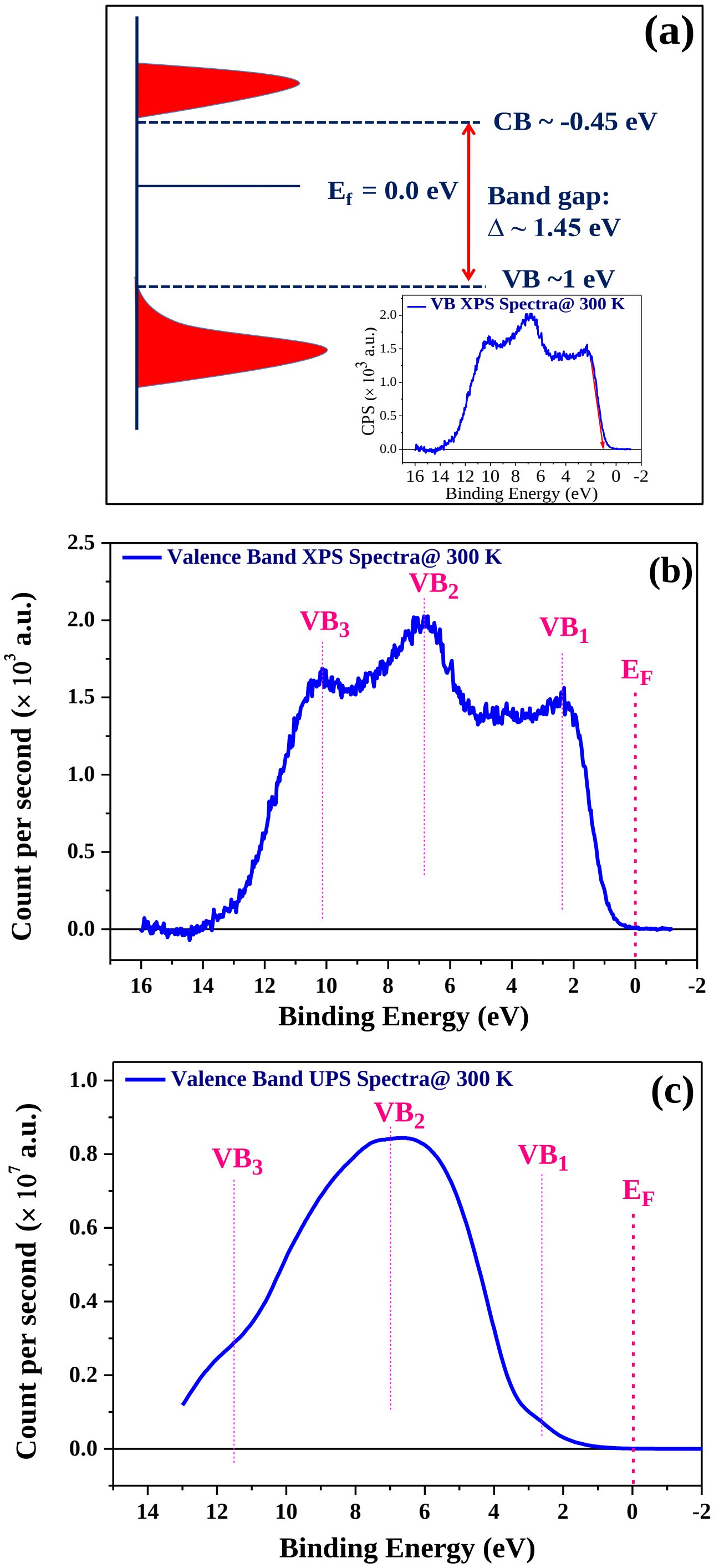

**Figure 5: (a)** Schematic diagram of the band structure. Inset shows the XPAS valence spectra of the composite0.7BFO-0.3TMO. The red line is guide to the eye showing the position of the edge of the valence band. **(b)** Valence band XPS spectrum of the composite at 300 K. **(c)** UPS spectrum of the composite at room temperature.

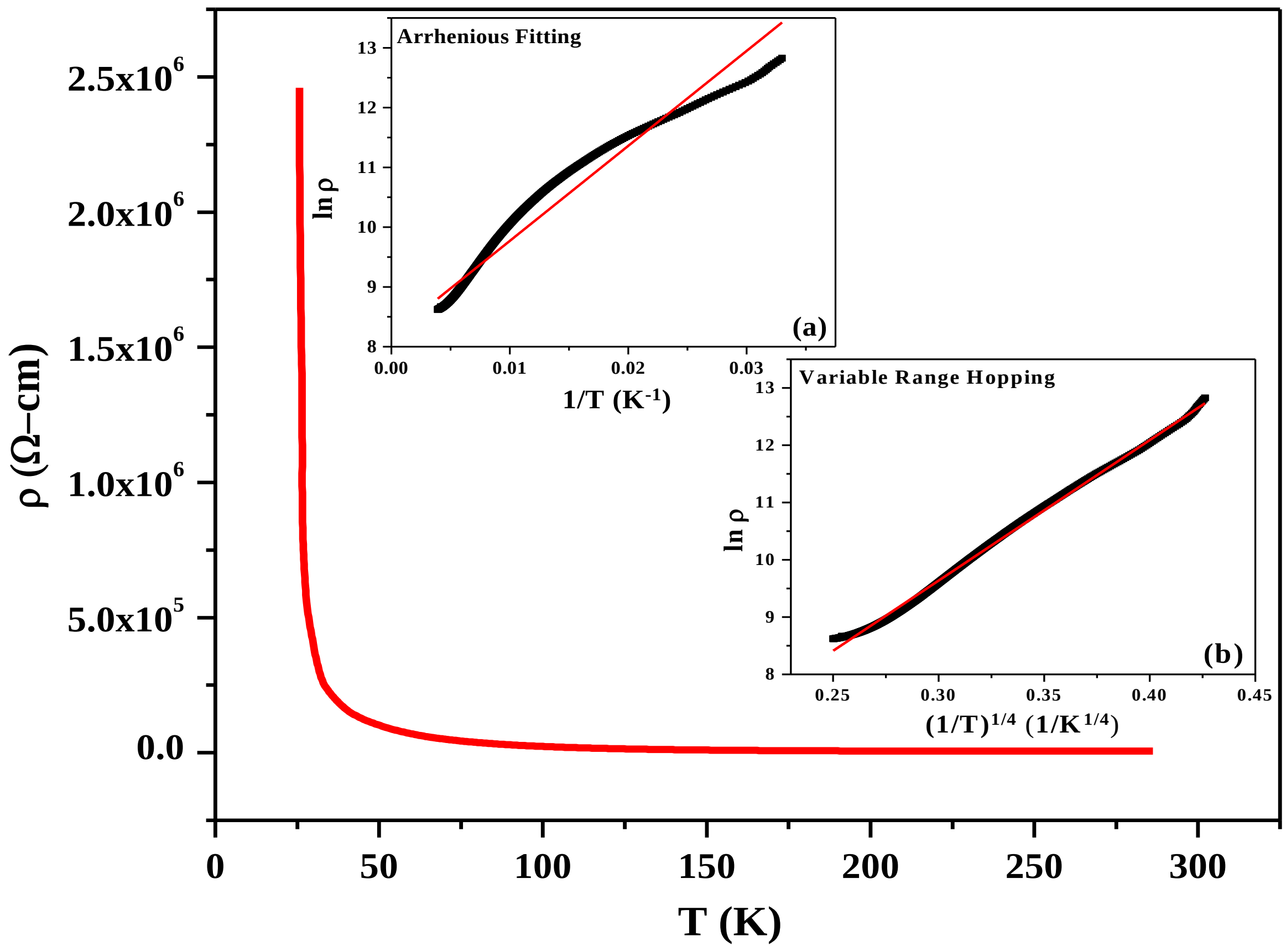

**Fig. 6:** Temperature dependent resistivity of the composite0.7BFO-0.3TMO. Inset (a) lnρ vs. 1/T plot showing Arrhenius plot. Inset (b) lnρ Vs. $(1/T)^{1/4}$ showing the plot for variable range hopping.

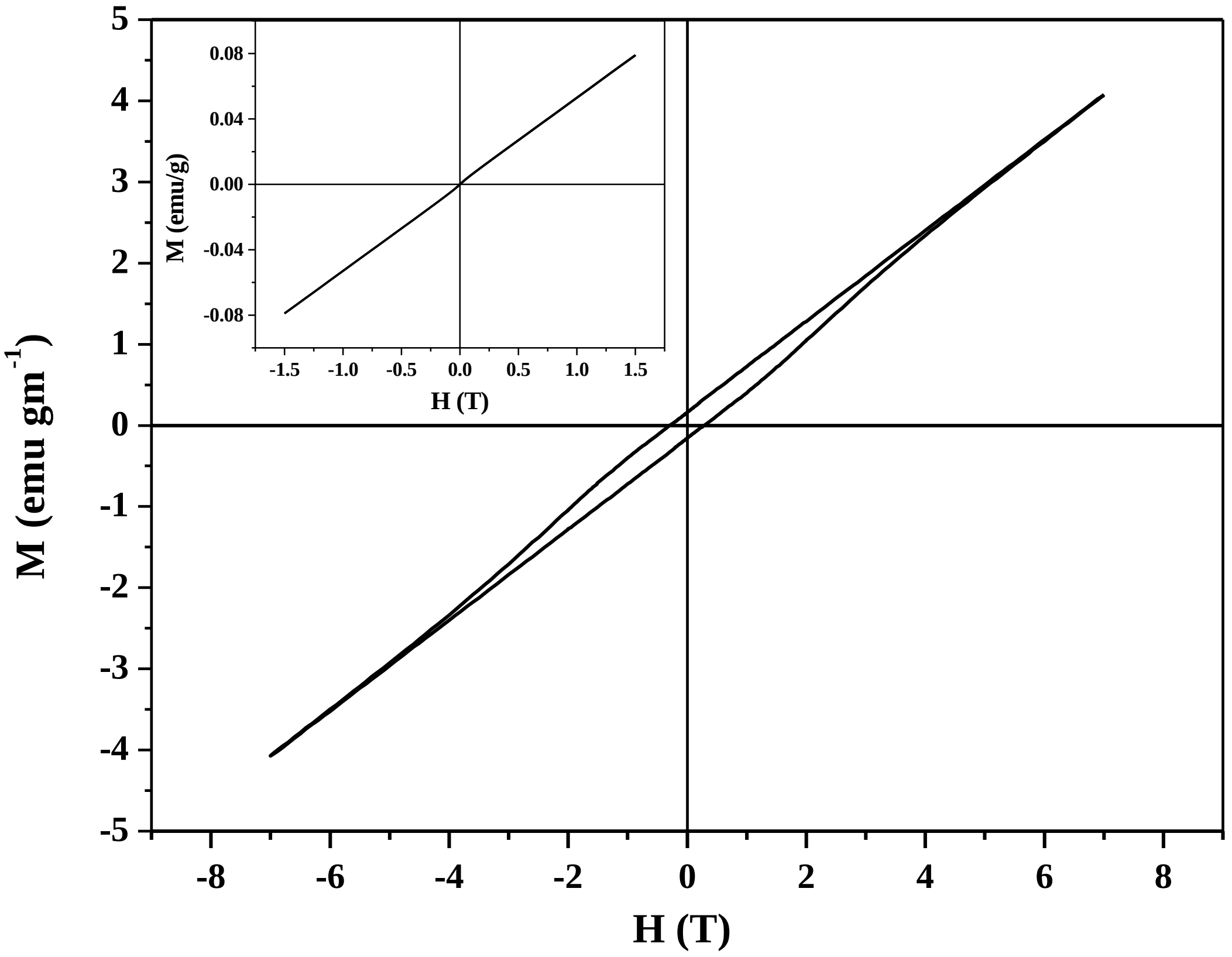

**Figure 7**: Room temperature M-H loop of the composite 0.7BFO-0.3TMO. Inset shows the room temperature M-H loop of pure $BiFeO_3$